\documentclass[conference]{IEEEtran}
\IEEEoverridecommandlockouts
\usepackage{cite}
\usepackage{amsmath,amssymb,amsfonts}
\usepackage{algorithmic}
\usepackage{graphicx}
\usepackage{textcomp}
\usepackage{xcolor}
\usepackage{subfig}
\usepackage{scalerel} %dee
\usepackage{tikz}
\usepackage{pgfplots}
\usetikzlibrary{arrows}
\usepackage{algorithm}
\usepackage{algorithmic}
\usetikzlibrary{shapes}
\usepackage{tabularx}

\pgfplotsset{width=2cm,compat=1.8}
\usetikzlibrary{svg.path}
\usetikzlibrary{patterns}
\def\BibTeX{{\rm B\kern-.05em{\sc i\kern-.025em b}\kern-.08em
    T\kern-.1667em\lower.7ex\hbox{E}\kern-.125emX}}
    \definecolor{orcidlogocol}{HTML}{A6CE39}
\tikzset{
  orcidlogo/.pic={
    \fill[orcidlogocol] svg{M256,128c0,70.7-57.3,128-128,128C57.3,256,0,198.7,0,128C0,57.3,57.3,0,128,0C198.7,0,256,57.3,256,128z};
    \fill[white] svg{M86.3,186.2H70.9V79.1h15.4v48.4V186.2z}
                 svg{M108.9,79.1h41.6c39.6,0,57,28.3,57,53.6c0,27.5-21.5,53.6-56.8,53.6h-41.8V79.1z M124.3,172.4h24.5c34.9,0,42.9-26.5,42.9-39.7c0-21.5-13.7-39.7-43.7-39.7h-23.7V172.4z}
                 svg{M88.7,56.8c0,5.5-4.5,10.1-10.1,10.1c-5.6,0-10.1-4.6-10.1-10.1c0-5.6,4.5-10.1,10.1-10.1C84.2,46.7,88.7,51.3,88.7,56.8z};
  }
}

\newcommand\orcidicon[1]{\href{https://orcid.org/#1}{\mbox{\scalerel*{
\begin{tikzpicture}[yscale=-1,transform shape]
\pic{orcidlogo};
\end{tikzpicture}
}{|}}}}

\usepackage{hyperref}

\begin{document}

\title{A Conflict-Aware Resource Management Framework for the Computing Continuum}

% %%%%%% Comment below for DOUBLE-BLIND REVIEW PROCESS
\author{
  \IEEEauthorblockN{Anonymous Author(s)}
  \IEEEauthorblockA{Anonymous Affiliation(s)}
}
\author{
   \IEEEauthorblockN{ 
   Vlad Popescu-Vifor\IEEEauthorrefmark{1}, Ilir Murturi\IEEEauthorrefmark{1}\IEEEauthorrefmark{4},  
   Praveen Kumar Donta\IEEEauthorrefmark{3},
    Schahram Dustdar\IEEEauthorrefmark{1}\IEEEauthorrefmark{2}
    }    
    \vspace{2mm}

    \IEEEauthorblockA{\IEEEauthorrefmark{1}Distributed Systems Group, TU Wien, Vienna, Austria
   } 
   \IEEEauthorblockA{\IEEEauthorrefmark{2}ICREA, Barcelona 08002, Spain   }
   \ \IEEEauthorblockA{\IEEEauthorrefmark{4}University of Prishtina, Department of Mechatronics, Prishtina, 10000, Kosova }
    \IEEEauthorblockA{\IEEEauthorrefmark{3}Department of Computer and Systems Sciences, Stockholm University, Stockholm 16425, Sweden
   }
    
    \vspace{-1cm}
}
 \IEEEoverridecommandlockouts
\IEEEpubid{\makebox[\columnwidth]{\hfill} \hspace{\columnsep}\makebox[\columnwidth]{ }}
%%%%%%%%%%%
\maketitle
\IEEEpubidadjcol
\begin{abstract}
The increasing device heterogeneity and decentralization requirements in the computing continuum (i.e., spanning edge, fog, and cloud) introduce new challenges in resource orchestration. In such environments, agents are often responsible for optimizing resource usage across deployed services. However, agent decisions can lead to persistent conflict loops, inefficient resource utilization, and degraded service performance. To overcome such challenges, we propose a novel framework for adaptive conflict resolution in resource-oriented orchestration using a Deep Reinforcement Learning (DRL) approach. The framework enables handling resource conflicts across deployments and integrates a DRL model trained to mediate such conflicts based on real-time performance feedback and historical state information. The framework has been prototyped and validated on a Kubernetes-based testbed, illustrating its methodological feasibility and architectural resilience. Preliminary results show that the framework achieves efficient resource reallocation and adaptive learning in dynamic scenarios, thus providing a scalable and resilient solution for conflict-aware orchestration in the computing continuum.

\end{abstract}

\begin{IEEEkeywords}
Computing Continuum, Resource Management, DRL, Orchestration
\end{IEEEkeywords}

\section{Introduction}
The growing complexity and heterogeneity of modern distributed systems have accelerated the shift toward the \textit{computing continuum}, a paradigm that seamlessly integrates cloud, fog, and edge resources to provide low-latency, context-aware computing capabilities. Driven by the rapid proliferation of connected devices and data-intensive applications (e.g., smart cities to real-time industrial automation), this integrated computing model enables optimizing computation across various tiers and ensuring adaptability and efficiency under varying workloads and resource constraints~\cite{pujol2023distributed, pujol2023edge}.

Orchestration frameworks such as Kubernetes have become essential for deploying and scaling services across computing continuum infrastructures. Mechanisms for managing dynamic resource conflicts in multi-agent, multi-objective optimization scenarios are still not well researched \cite{truyen2019comprehensive}. As deployments expand into increasingly heterogeneous and latency-sensitive environments, traditional resource management approaches, which are often static or purely reactive, fall short of meeting modern demands\cite{awad2022slo}. More specifically, computing infrastructures typically run a wide range of services, each with distinct performance requirements and resource usage patterns. These environments are inherently multi-agent \cite{soumplis2024performance, zhang2019heteroedge, chen2025agentflow }, where numerous components act concurrently to manage resources for different services. Resource conflicts arise when multiple agents issue contradictory actions. In this context, a conflict refers to incompatible resource specifications issued by distinct agents that cannot be simultaneously satisfied without degrading service performance or violating Service Level Objectives (SLOs). For example, during peak load, we might need to increase CPU allocation for critical services, while at the same time, another policy might aim to reduce infrastructure or device energy usage due to the high operating costs at a specific time. These competing goals can lead to resource conflicts, where multiple agents issue contradictory actions. One agent may attempt to reduce CPU usage to reduce costs, while another simultaneously requests more resources to maintain performance. These conflicting actions can create instability and cause repeated reconfigurations, degrade both system efficiency and service availability \cite{ naderializadeh2021resource}.

In this paper, we focus on a significant gap in computing continuum orchestration, which is the absence of automated and adaptive mechanisms for resolving resource conflicts arising from multiple agents. We introduce a framework that leverages Deep Reinforcement Learning (DRL) to detect, analyze, and resolve these conflicts in a scalable and self-adaptive manner. Through modeling the system state (i.e., including historical agent behaviors, resource utilization patterns, and performance indicators), the DRL can proactively suggest optimization strategies tailored to individual deployments and their operational thresholds. A key architectural feature of the proposed framework is its support for multiple, concurrently operating agents, each with distinct optimization goals such as minimizing energy consumption, reducing cost, or increasing computational throughput. These agents are modular and defined as specification objects, allowing for extensibility across various resource dimensions (e.g., CPU, memory, storage, etc.). 

Our core contributions within this paper are as follows:

\begin{itemize}
    \item A conflict-aware orchestration framework that integrates DRL with a Kubernetes-based environment, enabling adaptive and automated resolution of resource allocation conflicts.
    \item A synthetic simulation environment enabling controlled and repeatable evaluation of enforcement mechanisms under overlapping agent specifications that may cause resource conflicts.
    \item A feedback-driven reinforcement learning model, which evolves over time by learning from the outcomes of past conflict resolutions, thereby improving its future decision-making capabilities.
    \item Performing a preliminary evaluation of the proposed framework with respect to performance, latency, and resource efficiency.
\end{itemize}

The rest of this paper is structured as follows: Section~\ref{sec:relatedwork} reviews related work in resource management and intelligent orchestration in distributed systems. Section~\ref{sec:framework} introduces the framework and outlines its main architectural components.  
Section~\ref{sec:evaluation} discusses implementation details and evaluates the proposed framework through preliminary experiments. Finally, Section~\ref{sec:conclusion} concludes with a summary of findings and directions for future research.

\section {Related Work}
\label{sec:relatedwork}

The computing continuum aims to create a dynamic and adaptive computing environment that spans between cloud and edge computing \cite{murturi2023learning, murturi2021decentralized}.  A critical point emphasized in this research is the inadequacy of the reactive management system for computing continuum systems due to the unique demands.% Such  

Li et al.\cite{li2025adaptive} introduce a graph-augmented multi-agent reinforcement learning system where local agents optimize resources in their neighborhoods, while a global orchestrator handles system-wide coordination. This collaborative MARL architecture effectively resolves competing goals among agents across heterogeneous infrastructure layers. However, the proposed approach is given as a high-level architecture without evaluation.   Several important contributions in this field utilize DL and RL techniques to address various challenges. Most of these efforts primarily focus on resource allocation and management; however, they often neglect the crucial aspect of reactive interventions when conflicts arise between agents' actions in the system.  A notable example is the DRL model proposed by Zeng et al. \cite{zeng2019resource}. This model offers a model-free solution designed to automatically predict user resource consumption patterns by analyzing historical interaction data. While this model presents a significant advantage through its model-free nature, it is constrained by its dependency on the specific environment in which it is deployed.  
Naderializadeh et al.  \cite{naderializadeh2021resource} suggest a DRL distributed model, which runs across multiple agents. This paper describes a more complex deep learning model that depends on the communication between the distributed agents. An important aspect of this paper is that the agents are independent and may make decisions simultaneously while being unaware of the other agent's decisions. An interesting DRL implementation is presented by Liu et al.\cite{liu2017hierarchical}, which uses Q-Learning to derive the optimal choices when facing specific situations. 
In contrast to existing research, which primarily focuses on resource allocation optimization or workload prediction in isolated domains of the computing continuum, this work introduces a conflict-aware orchestration mechanism that addresses coordination challenges arising from user-defined resource management specifications. The proposed approach integrates a feedback-driven, self-adaptive mechanism that continuously learns from system performance and previous optimization outcomes.

\section{The Framework}
\label{sec:framework}
 
\begin{figure}[h]
    \centering
    \includegraphics[width=\columnwidth]{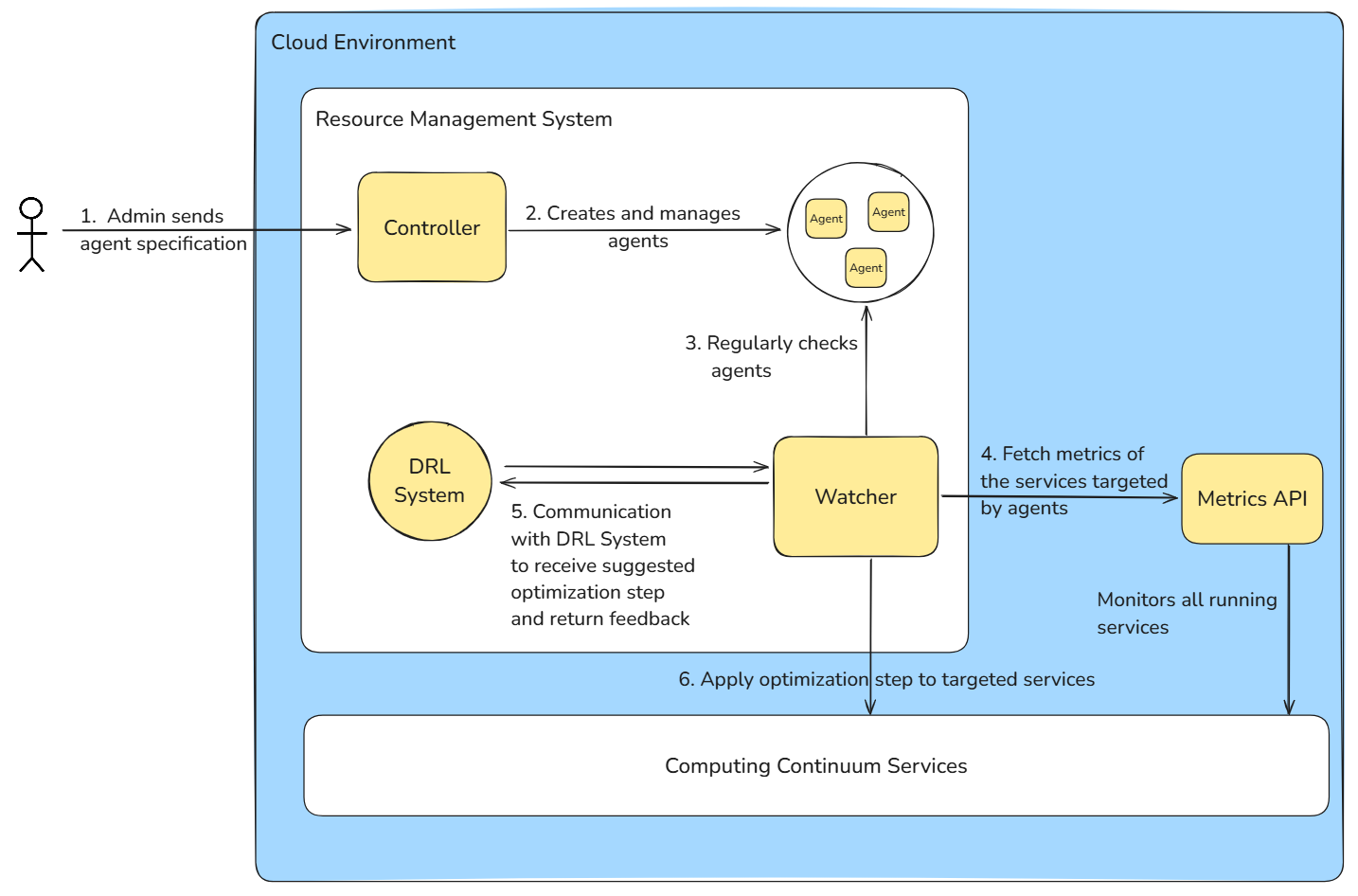}
    \caption{Workflow and communication of the framework.}
    \label{fig:framework}
\end{figure}

In Figure \ref{fig:framework}, we illustrate the main components of the proposed framework: (i) agents, (ii) the controller, (iii) the watcher, (iv) the DRL system, and (v) the metrics API.

\subsection{Agents}

An agent is a structured object that represents the resource specifications defined by the user and instantiated by the Controller in a desired cluster. An agent is stored in a cluster state to ensure that other components, such as the Watcher, can continuously access and update it as needed.   
An agent object follows a defined structure that allows other components in the system to interpret and act on the resource specifications.  User-defined specifications are expressed as declarative YAML/JSON objects that describe the desired resource management rules such as (i) scope (i.e., indicates the level at which the resource specification should be applied), (ii) status (i.e., state of the agent (e.g., pending, conflicting, etc.)), (iii) target (i.e., specifies the exact target of the resource specification), (iv) resource quota (i.e., CPU or memory limits to be applied), and (v) restart count (i.e., tracks the number of times the agent's specification has been reapplied). Furthermore, two categories of conflicts are identified during execution: (i) specification conflicts and (ii) optimization conflicts. Specification conflicts occur when multiple agents repeatedly override each other’s configurations, as detected by the watcher when a specification must be reapplied several times consecutively. Optimization conflicts arise when an applied configuration degrades performance, such as increased latency or memory exhaustion; these are resolved adaptively through the DRL system (as explained in III-D).

\subsection{The Controller}

The controller serves as the main entry point for resource management, and it is responsible for creating and managing agents based on incoming requests.  The controller receives structured HTTP requests from the end-user and, after validating the input, uses the provided specifications to create or update agent objects.  
Depending on the request, the controller creates, modifies, or removes an agent within a target cluster. A dedicated endpoint allows end-users to define detailed resource constraints at different scopes, such as nodes (i.e., applies to all workloads on a given node), namespaces (i.e., applies to all deployments within a namespace), and deployments (i.e., applies fine-grained optimization to a single deployment). A deployment represents a single application or service component managed by Kubernetes, typically defined by a container image, its replicas, and resource constraints.

\subsection{The Watcher}

 The watcher is responsible for monitoring both active and pending agents and acts as a coordinator to ensure that agent specifications are correctly applied to a cluster. The watcher connects to a cluster API to stay updated on agents and ensure new objects within the defined scope receive the specified configurations. In addition, the watcher monitors resource consumption of the services targeted by each agent and enforces the defined specifications when deviations are detected.  

The watcher operates iteratively, periodically verifying whether each agent’s specification is correctly enforced on its target. Once applied, the agent remains under continuous verification. If deviations are detected, the specification is reapplied to maintain consistency. When the same specification must be reapplied repeatedly, the agent is marked as conflicting, and a counter tracks the conflict duration. Such logical conflicts usually result from overlapping specifications and are resolved by the watcher; only unresolved cases are forwarded to the DRL system for adaptive handling.

 \subsection{Deep Reinforcement Learning (DRL)}
 The DRL system is invoked when runtime performance deteriorates following an applied specification or when logical conflicts remain unresolved by the watcher. In such cases, it adaptively selects corrective actions such as scaling, migration, or resource reallocation. The DRL system operates as a service hosting a neural network model trained to make optimization decisions for targeted deployments. Furthermore, the DRL system relies on real-time metrics such as CPU utilization, memory usage, and latency which are retrieved through the Metrics API. Two complementary mechanisms form the basis of the system. The general-purpose Meta DRL Model captures common optimization patterns and supports system-wide decision-making. When this meta model yields suboptimal outcomes for specific services, an Adaptive DRL mechanism via Instance-Based Learning retrains specialized models using feedback data. These service-specific models evolve over time, improving the precision of optimization actions. The following subsections describe the meta-level model for general optimization and the adaptive mechanism that enables reinforcement learning at the instance level \cite{guo2020learning}.

\subsubsection{Meta DRL Model for General Optimization}

The input structure for training the model is based on deployment-specific data such as the number of replicas in the deployment, average CPU/memory usage, CPU/memory allocation, and response latency (i.e., before and after applying an agent specification). These input features enable the model to evaluate differences in resource usage and latency. The watcher component supplies real-time memory and latency data by connecting to a monitoring system.  
The optimization decision returned by the model is represented as an integer, mapped as follows: \texttt{0: Optimize (no changes needed)}, \texttt{1: Migrate to another node}, \texttt{2: Scale down}, \texttt{3: Scale up}. These actions are triggered based on specific conditions. For example, a pod may be migrated to another node if latency increases significantly, if it experiences memory exhaustion (e.g., OOMKill), or if memory allocation is insufficient for stable operation.  

\subsubsection{Adaptive DRL via Instance-Based Learning}

\begin{figure}[h]
    \centering
    \includegraphics[width=\columnwidth]{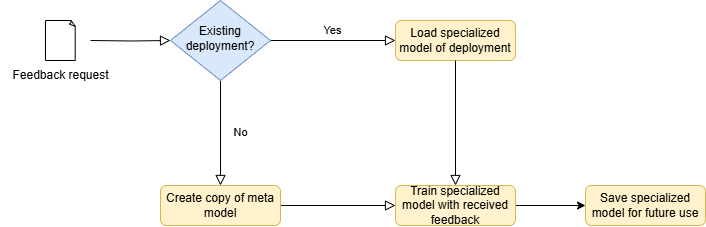}
   \caption{Flow diagram of DRL feedback process.}
    \label{fig:flow-feedback}
\end{figure}

An instance-based reinforcement learning approach is integrated into the DRL system. This mechanism builds upon the main DRL model, referred to as the meta model, which is designed to operate with common threshold values applicable to most deployments. When the system receives negative feedback regarding an optimization decision, it responds by cloning the meta model to create a deployment-specific version. This specialized model is retrained using the corrective feedback and is subsequently reused for future optimization tasks targeting the same deployment. Figure~\ref{fig:flow-feedback} illustrates this decision flow.

The DRL method maintains a registry of specialized models using a dictionary structure within the system container. When an optimization request is received, the method first checks if a model specifically designed for the given deployment already exists. If such a model is available, it is loaded to generate a prediction. If not, a new model is created by cloning the meta model and training it with feedback data that includes the suggested corrective action and its associated reward score. This specialized model is then stored in a persistent volume to ensure continuity across system restarts. Furthermore, our approach facilitates the progressive development of deployment-specific models, and over time, these models refine their predictions, particularly in cases where services are frequently re-evaluated. % 

\begin{figure}[h]
    \centering
    \includegraphics[width=\columnwidth]{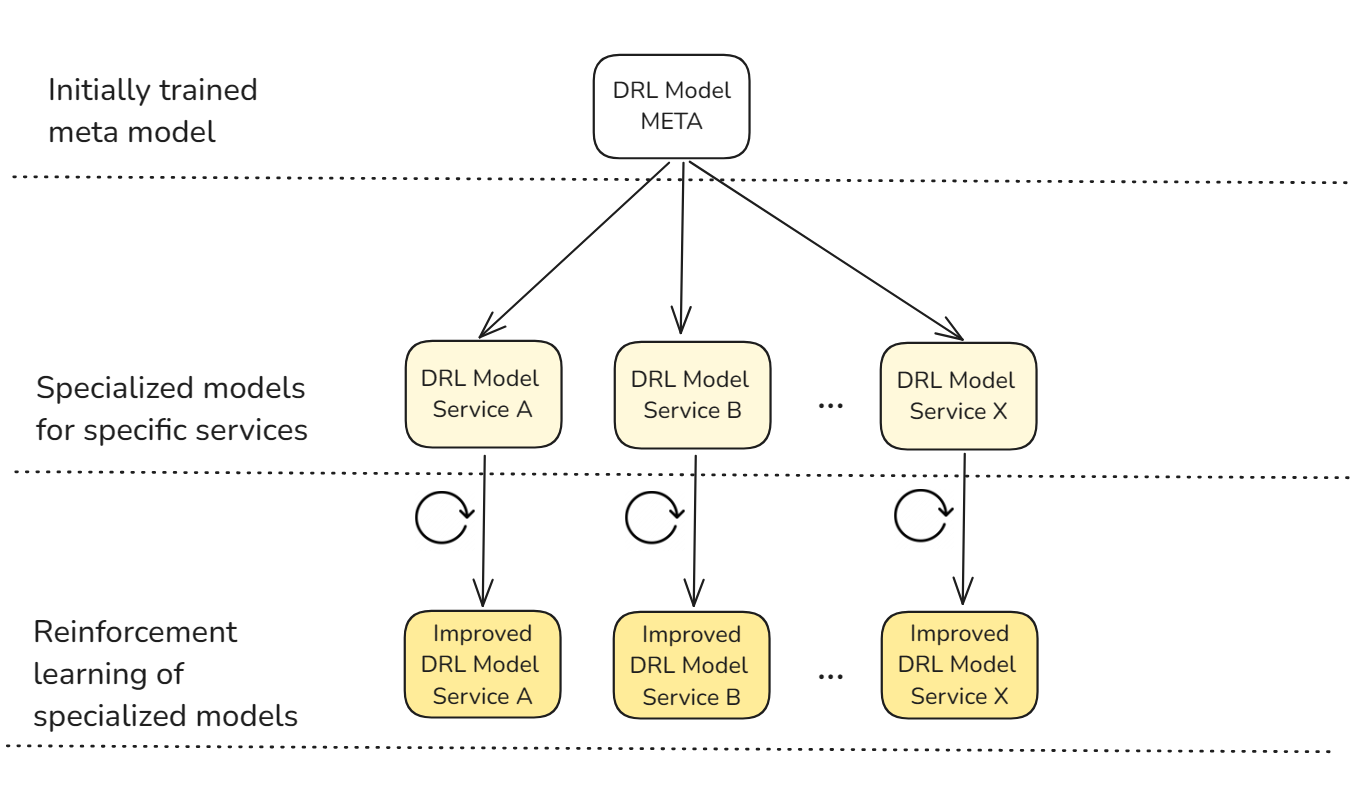}
    \caption{An overview of the DRL system.}
    \label{fig:drl-system}
\end{figure}

After an agent specification is registered, the watcher gathers metrics and forwards them to the DRL system for optimization inference.  
Figure~\ref{fig:drl-system} illustrates the overall DRL process and the interaction between the meta model and its adaptive variants.

The watcher maintains a continuous connection to the DRL system, facilitating iterative optimization. The watcher evaluates conflicting agents, retrieves recommendations from the DRL model, and implements these actions while monitoring service behavior. If performance issues, such as increased latency or memory errors, occur, feedback is sent to refine the model. Such a process enables the DRL system to improve decision-making and adapt to different cloud environments for better optimization.

\section{Implementation and Evaluation}
\label{sec:evaluation}
This section presents the testbed\footnote{DRL-RMS, https://github.com/vladfreeze/DRL-RMS-Framework} and numerical evaluation of the proposed DRL-based conflict-aware resource management framework. Even though the evaluation is preliminary, the contribution of this work is primarily methodological and architectural, rather than purely performance-based. Note that the current evaluation focuses on validating the initial part of the framework, specifically node-level optimization and watcher enforcement under conflict-prone conditions. Experimental validation of multi-agent conflict resolution and DRL-based adaptive optimization is reserved for future work.

\subsection{Testbed Setup}
A Kubernetes-based testbed was deployed on a virtualized infrastructure (i.e., simulating a computing continuum). For simplicity, the testbed is deployed as a single cluster; however, the framework is designed to seamlessly support multi-cluster and distributed configurations. The primary reason to use Kubernetes due to its maturity, modularity, and extensive support for container orchestration. To ensure flexibility and scalability, the cluster nodes were provisioned using \textit{Harvester}, an open-source hypervisor developed by SUSE\footnote{Harvester: Open Source HCI Solution}. Each virtual machine (VM) runs Ubuntu Server 24.04 with about 300 GB of storage. IP addresses are dynamically assigned via DHCP within a cluster subnet. The control plane nodes each have 8 virtual CPUs (vCPUs) and 32 GB of RAM for managing the Kubernetes cluster and resource management framework, while the worker nodes are equipped with 4 vCPUs and 8 GB of RAM to simulate edge environments. The cluster comprises three control plane nodes and four worker nodes, which allows for testing horizontal and vertical scaling and dynamic conflict resolution due to competing resource allocation policies.  

The cluster service architecture is based on RKE2\footnote{RKE2: \url{https://docs.rke2.io}}, a lightweight Kubernetes distribution known for its easy deployment and ARM compatibility. Management is performed using SUSE Rancher\footnote{SUSE Rancher: \url{https://www.rancher.com}}, which offers an intuitive interface for Kubernetes configuration. The cluster uses MetalLB\footnote{MetalLB: \url{https://metallb.universe.tf}} for external IP addressing in bare metal environments. Monitoring is provided by Prometheus\footnote{Prometheus: \url{https://prometheus.io}} and Grafana\footnote{Grafana: \url{https://grafana.com}}, which track performance metrics. A private Docker registry supports container deployment, and Jupyter Lab\footnote{Jupyter Lab: \url{https://jupyter.org}} is deployed for interactive development and testing of components like the Watcher and DRL system.
Resource consumer services run on worker nodes, supported by the orchestration layer. These containerized services mimic applications needing dynamic resource optimization. They train a small machine learning model and log execution times to Prometheus for real-time monitoring and optimization without disrupting workflows. 

%train a small machine learning model
\begin{figure}[t]
\centering
\includegraphics[width=\columnwidth]{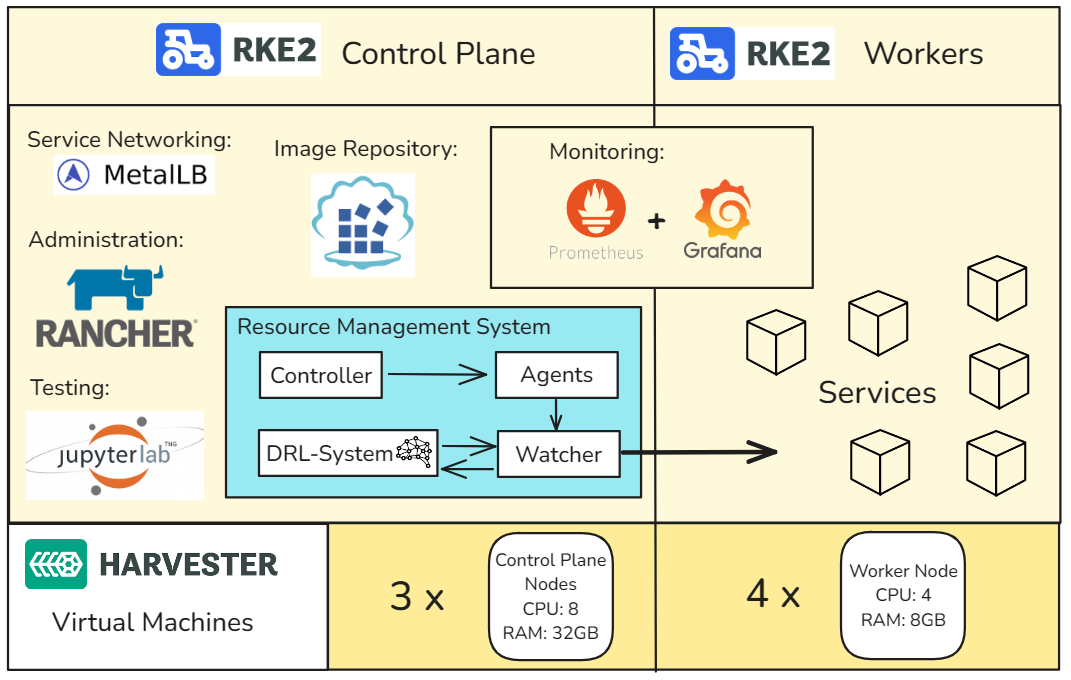}
\caption{Cluster architecture diagram.}
\label{fig:architecture}
\end{figure}

\subsection{Preliminary Evaluation: Node-Level Optimization}
The following experiment evaluates the proposed framework's behavior under a conflict-prone condition where a single node experiences a resource imbalance that could lead to contention between agents. The cluster was configured so that one node, specifically \textit{worker-1}, hosted three services while the other nodes hosted only two, resulting in a CPU utilization of approximately 75\% compared to around 50\% on the remaining nodes. To mitigate the imbalance, a node-scoped agent was deployed by the controller targeting \texttt{worker-1} with a CPU quota of 0.85, signaling a desired 15\% reduction in resource consumption. After the agent was instantiated, the Watcher detected the new specification and initiated the enforcement process. The agent accessed the corresponding deployments on the target node, retrieved performance metrics from Prometheus, and applied new CPU limits by reducing observed usage by about 30\%. Throughout the enforcement cycle, no specification conflicts were triggered. Meaning that, the Watcher maintained consistent configurations without repeated reapplications, and no optimization conflicts were observed, as service performance remained stable. Table~\ref{tab:bench1-lat} summarizes the results, showing reduced CPU allocations accompanied by a modest latency increase for the services hosted on \textit{worker-1}.

\begin{table}[ht]
\small
\centering
\caption{Optimization results on \textit{worker-1}.}
\label{tab:bench1-lat}
\begin{tabular}{|c|c|c|c|c|}
\hline
\textbf{Service} & \textbf{CPU$_\text{init}$} & \textbf{CPU$_\text{opt}$} & \textbf{Lat.$_\text{init}$ (s)} & \textbf{Lat.$_\text{opt}$ (s)} \\
\hline
cons-a & 0.94 & 0.80 & 12.6 & 14.9 \\
cons-b & 0.92 & 0.78 & 12.2 & 15.1 \\
cons-c & 1.00 & 0.85 & 12.1 & 14.3 \\
\hline
\end{tabular}
\end{table}

As shown in the Table~\ref{tab:bench1-lat}, latency increased by approximately 18\% on average, but all values remained within performance thresholds that had been previously identified during the model training phase. These changes confirmed that the optimization was successful and did not compromise service responsiveness beyond acceptable limits. Table~\ref{tab:b1-node-results} further summarizes the resource utilization metrics observed at the node level. CPU usage on worker-1 decreased from 75\% to 60.75\%, confirming that the agent’s target and updated specifications were successfully enforced. 

\begin{table}[ht]
\small
\centering
\caption{Node-level resource usage after optimization.}
\label{tab:b1-node-results}
\begin{tabular}{|c|c|c|c|}
\hline
\textbf{Node} & \textbf{CPU Use (\%)} & \textbf{CPU Res. (\%)} & \textbf{Mem Res. (\%)} \\
\hline
worker-1 & 60.75 & 60.75 & 56 \\
\hline
\end{tabular}
\end{table}

\begin{figure}[h]
    \centering
    \includegraphics[width=\columnwidth]{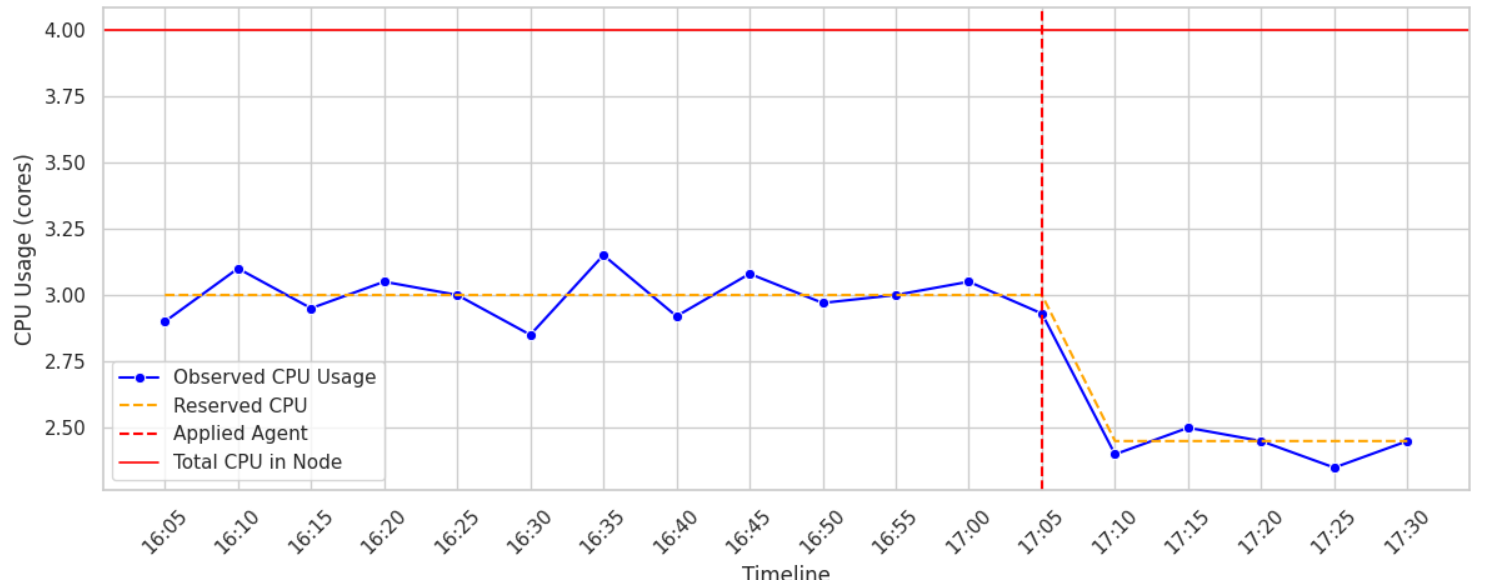}
    \caption{CPU usage on \textit{worker-1} after optimization.}
    \label{fig:benchmark-cpu}
\end{figure}

The timeline of CPU usage is shown in Figure~\ref{fig:benchmark-cpu}. The quota was enforced at 17:05, and the immediate drop in CPU utilization demonstrates the system’s responsiveness and ability to apply policy changes dynamically.

\section{Conclusion}
\label{sec:conclusion}
This paper introduces a conflict-aware resource orchestration framework using DRL to manage agent-level resource conflicts in computing continuum environments. The implementation on a virtualized cluster showed that applying the framework reduced node-level CPU usage from 75\% to 60.75\% and optimized service-level resource quotas by 15-20\%. Our initial evaluation of conflict-aware resource management showed promising results from controlled simulations. However, it may not reflect real-world complexities, and areas like scalability and long-term adaptability, which need further investigation. Future work will extend the framework toward cluster-level optimization and include experimental demonstrations of agent-level conflict detection and resolution under realistic workloads.

\section*{{Acknowledgment}}
Research has partially received funding from a grant agreement No. 101135576 (INTEND).

\bibliographystyle{ieeetr}
\bibliography{sample-base} 
\end{document}